\documentclass[preprint, review,3p,times, 10pt]{elsarticle}

\usepackage{cite}
\usepackage{amsmath,amssymb,amsfonts}
\usepackage{algorithmic}
\usepackage{graphicx}
\usepackage{textcomp}
\usepackage{comment}
\usepackage{footnote}
\usepackage{comment}
\usepackage{threeparttable}
\usepackage{romannum}
\usepackage{longtable}
\usepackage{lscape}
\usepackage{soul}
\usepackage{array, makecell}
\usepackage{multirow}
\usepackage{enumitem}
\usepackage{tikz}
\usepackage{pgfplots}
\usepackage{pgf-pie}
\usepackage{longtable}
\usepackage{subcaption}

\usepackage[hyphens]{url}
\usepackage{hyperref}
\hypersetup{colorlinks=true,breaklinks=true}

\usepackage{pgfplots}
\usepgfplotslibrary{statistics}
\pgfplotsset{compat=1.18}

\journal{'}

\begin{document}

\begin{frontmatter}
\title{Advancing Web Browser Forensics: Critical Evaluation of Emerging Tools and Techniques}

\author[1]{Rishal Ravikesh Chand}\ead{rishalc@unifiji.ac.fj}
\author[1]{Neeraj Anand Sharma}\ead{neerajs@unifiji.ac.fj}
\author[2]{Muhammad Ashad Kabir \corref{correspondingauthor}}
\cortext[correspondingauthor]{Corresponding author: Charles Sturt University, Panorama Ave, Bathurst, NSW 2795. Ph.+61263386259}
\ead{akabir@csu.edu.au}%
\affiliation[1]{organization={Department of Computer Science and Mathematics, The University of Fiji}, city={Lautoka}, country={Fiji}}
\affiliation[2]{organization={School of Computing, Mathematics and Engineering, Charles Sturt University}, city={Bathurst}, state={NSW}, postcode={2795}, country={Australia}}

\begin{abstract}

As the use of web browsers continues to grow, the potential for cybercrime and web-related criminal activities also increases. Digital forensic investigators must understand how different browsers function and the critical areas to consider during web forensic analysis. Web forensics, a subfield of digital forensics, involves collecting and analyzing browser artifacts, such as browser history, search keywords, and downloads, which serve as potential evidence. While existing research has provided valuable insights, many studies focus on individual browsing modes or limited forensic scenarios, leaving gaps in understanding the full scope of data retention and recovery across different modes and browsers. This paper addresses these gaps by defining four browsing scenarios and critically analyzing browser artifacts across normal, private, and portable modes using various forensic tools. We define four browsing scenarios to perform a comprehensive evaluation of popular browsers -- Google Chrome, Mozilla Firefox, Brave, Tor, and Microsoft Edge -- by monitoring changes in key data storage areas such as cache files, cookies, browsing history, and local storage across different browsing modes. Overall, this paper contributes to a deeper understanding of browser forensic analysis and identifies key areas for enhancing privacy protection and forensic methodologies.
\end{abstract}

\end{frontmatter}

\section{Introduction}
\label{sec:introduction}

Cybercrime has become increasingly prevalent as digital technology and the internet have evolved. Criminals exploit vulnerabilities in computer systems and use the internet for malicious purposes \citep{1}. One significant avenue for cybercrime is through web browsers, which are widely used to access websites, stream multimedia, maintain social profiles, and more \citep{1}. Given the widespread use of browsers, the exponential growth of internet resources and technology has unfortunately led to increased web-related criminal activities. Browsers store a wide range of artifacts such as browsing history, search keywords, downloads, URLs, user autofill details, forms activity, streaming and search information, and saved logins, which can be critical in reconstructing online activities \citep{2,3}.

Browser artifacts are crucial for digital forensic investigations as they provide valuable evidence of a user's actions and interactions within a browser. These artifacts can help investigators trace the sequence of events, recover deleted data, and establish timelines, making them essential investigations. While previous studies \citep{4,5,6} have outlined the basic procedures for browser forensics, they have not sufficiently explored effective investigative approaches across all three browsing modes -- normal, private, and portable. This lack of focus limits the evaluation of the artifacts recovered and hinders conclusions regarding the effectiveness of each browsing mode.

This paper aims to bridge the gaps in the literature by providing a comprehensive exploration of browser artifacts and applying advanced forensic tools to analyze and extract valuable data. By examining browser forensics, this study enhances the investigator's capabilities in tackling evolving cybercrime. Our approach includes a thorough analysis of browser artifacts, combined with the application of sophisticated tools to improve data extraction and analysis techniques. Our study covered Google Chrome, Mozilla Firefox, Brave, Tor, and Microsoft Edge, where we performed typical web activities such as browsing, keyword searches, and downloading web content. Through planned scenarios, we assessed how each browser handled and retained user information after activities like page visits, downloads, and logins. The results revealed that despite the browser's privacy claims, artifacts like search keywords, URLs, and session details were recoverable even after clearing history and cache. This detailed examination is essential for forensic investigations, as it highlights the persistent nature of data across different browsing environments, providing valuable insights into the effectiveness of privacy measures and the potential for forensic recovery. This evaluation gives forensic investigators precise insights into which browsers and modes will likely retain the most valuable data for digital forensic investigations.
The findings of this study offer digital forensic investigators practical methodologies and tools for recovering artifacts from devices and browsers.
This research makes the following key contributions:
\begin{itemize}
    \item We summarize the latest techniques and tools used in browser forensics, highlighting their effectiveness in recovering crucial artifacts from various devices and browsers.
    \item We define four browsing scenarios to conduct a critical analysis of browser artifacts across different browsing modes -- normal, private, and portable -- using various forensic tools. 
    \item We conduct a comprehensive evaluation of popular browsers, Google Chrome, Mozilla Firefox, Brave, Tor, and Microsoft Edge, by monitoring changes in key data storage areas such as cache files, cookies, browsing history, and local storage across different browsing modes. 
\end{itemize}

The rest of the paper is organized as follows. Section~\ref{sec:background} outlines the research background of browser forensics and the techniques. Section~\ref{sec:related-work} reviews related work, highlighting different approaches, and artifacts that can be recovered, and identifies limitations and research gaps in the current literature. Section~\ref{sec:methodology} summarizes the methodology used in the study, describing the environment setup for testing, the tools utilized, and the data acquisition and analysis processes for normal, private, and portable browsing modes. Section~\ref{sec:analysis} provides an analysis of the recoverable artifacts identified using various tools across different browsing modes, followed by the presentation of results in Section~\ref{sec:results}. Section~\ref{sec:discussion} offers a detailed discussion, comparing the various types of artifacts recovered in each browsing mode and addressing the research questions. Section~\ref{sec:challenges} discusses the challenges encountered and provides recommendations for future work. Section~\ref{sec:conclusion} concludes the paper, by summarizing the key findings and their implications.

\section{Background}\label{sec:background}


This section provides an overview of web browser architecture and functionality, as well as browser forensics techniques. These include the analysis of browsing history, cache, cookies, bookmarks, downloads, passwords, malware, live memory, and cloud-based extensions. These methods are vital for extracting digital evidence in forensic investigations.

\subsection{Overview of Browsers}
A web browser is an application software primarily used to access the internet and online resources. It provides users with various types of user interfaces, allowing them to navigate websites and retrieve web pages, images, text, and load different types of data. Users can access these resources via desktop or mobile browsers. Requests are sent from the browser to the web servers, and information is securely exchanged using the Hypertext Transfer Protocol Secure (HTTPS)~\citep{12}. The data is then retrieved by the web browser and displayed to the user. 
Mobile browsers operate on handheld devices and are responsible for displaying web pages on these smart handheld devices~\citep{13,14}. When a user requests a web page via a URL, the browser retrieves and parses the HTML into the appropriate structure for display on the mobile screen and renders web content, such as images and text, and executes JavaScript, and handles user input~\citep{15,16}.  

\subsection{Web Browser Architecture}
\begin{figure}[!htb]
    \centering
    \includegraphics[width=0.6\linewidth]{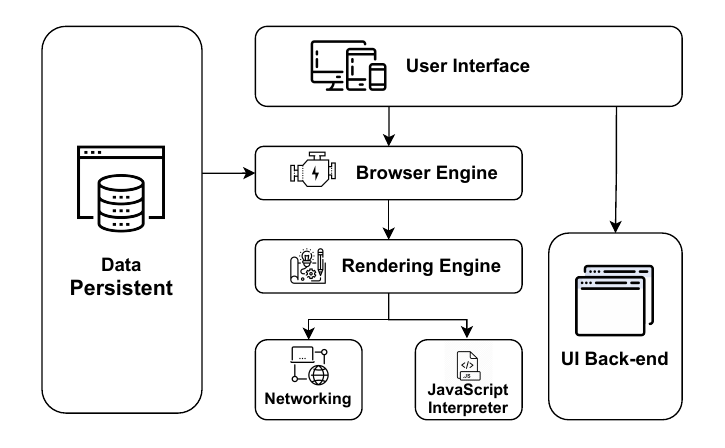}
    \caption{Browser architecture}
    \label{fig:1}
\end{figure}

Web browsers have become essential tools in modern life, providing users with quick access to news, media, and current affairs. Browsers follow a defined architecture that facilitates user interaction with the internet. This architecture comprises two key components: the front-end and the back-end. The front-end is designed to be user-friendly, ensuring easy of use, while the back-end manages the core functions of the browser~\citep{17,18}, as illustrated in Figure~\ref{fig:1}. 

\paragraph{User Interface} It refers to the visual and graphical elements displayed to the user. It allows user to interact with visual components, such as buttons, labels, menus, and clickable icons.

\paragraph{Rendering Engine} The rendering engine is responsible for interpreting and displaying the content of web pages. It serves as a bridge between the graphical user interface (GUI) and the web page's underlying code, rendering the website's style sheets and other visual effects. Each web browser uses a different rendering engine~\citep{12,19}, as reported in Table~\ref{tab:2}. 

\newpage
\begin{table}[ht!]
\centering
\caption{Browsers rendering engines}
\begin{tabular}{ll}
\hline
Browser & Render Engine \\
\hline\hline
Google Chrome & Blink \citep{20} \\

Mozilla Firefox & Gecko \citep{21} \\

TOR Browser & Gecko \citep{21} \\

Internet Explorer & Trident \citep{22} \\

Chrome (iOS) & WebKit \citep{23} \\

Safari (iOS) & WebKit \citep{23} \\
\hline

\hline
\end{tabular}
\label{tab:2}
\end{table}

\paragraph{Browser engine} It connects the GUI with the rendering engine, facilitating interactions based on the user input and manipulation of the interface. 

\paragraph{Networking} It is responsible for establishing connections between the web browser and web servers. It manages security certificates and handles other various internet protocols such as file transfer protocol (FTP). 

\paragraph{JavaScript interpreter} It is responsible for executing JavaScript code on web pages, enabling dynamic content and interactive features.

\subsection{Overview of Browser Forensics}

Digital forensics encompasses the identification and collection of critical evidence through various techniques and tools, all conducted in an orderly manner with proper documentation as mandated by law. This field involves the search and custody of electronic devices, networks, media, and gadgets, as well as the extraction of evidence that contributes to forensic investigations \citep{2}. Digital forensics is often considered more of an art than a science, as it requires specialized expertise in various techniques, including email, database, malware, disk, and browser forensics.

Existing research in browser forensics focuses on analyzing and retrieving stored data from various web browsers, including Mozilla Firefox, Google Chrome, Internet Explorer, and Safari. Moreover, there is a growing emphasis on conducting a structural analysis of internet log files to uncover traces of online activity from a forensic perspective. The primary objective is to gather evidence regarding the online behaviors of the subject under investigation.

Searching for evidence related to web browsing activity is often essential in digital forensic investigations. Virtually all actions taken by a suspect while using a web browser can leave a trail on their computer, including even basic searches \citep{24}. Therefore, examining this evidence on a suspect's device can provide valuable information for investigators. The retrieval of data such as cache, history, cookies, and download lists can facilitate the analysis of visited web pages, as well as the frequency and duration of access, and the search terms employed by the suspect \citep{25}. Various techniques are available in web browser forensics to effectively retrieve this evidence.

\subsection{Browser Forensics Techniques}
\paragraph{Browser history analysis} In this technique, the browser history is used to verify the user's digital footprints.
This technique analyzes which sites and web pages were accessed by the suspect, as recorded in the history logs, which typically include the date, time, and frequency of visits to each webpage. This method is critical in digital browser forensics, as it provides compelling evidence of the suspect's online activities. It also assists investigators in understanding the suspect's behavior and interests during their internet usage.

To conduct browser history analysis, investigators utilize specialized forensic applications to acquire data from the suspect's confiscated device. This data retrieval provides valuable insights into the suspect's online activities and behavior, helping to establish a comprehensive timeline of events \citep{26}. During the acquisition process, various data anomalies may be discovered, allowing investigators to identify patterns in the suspect's behavior, which can further aid the investigation. 
    
\paragraph{Cache analysis} Cache files contain fragments of information such as images, text files, scripts, and other related data saved on the device used to access websites \citep{27}. Analyzing these files, commonly referred to as cache analysis, involves examining the contents and various types of files stored during the suspect's browsing sessions. This technique is useful as it provides a comprehensive overview of the browser's cache, allowing investigators to access crucial information that can assist in the investigation \citep{28}.

\paragraph{Cookies analysis} Cookies are small text files created by web browsers that store user preferences, login credentials, and information related to the web pages visited by the user \citep{29}. In a digital forensics investigation, analyzing cookies can provide critical insights into user behavior, preferences, sites visited, and stored credentials, all of which can serve as significant evidence. To conduct cookies analysis, investigators must acquire the suspect's device and use specialized software to analyze and extract cookies associated with the web browser in use. Overall, cookie analysis is a vital technique in digital forensics, as it can uncover valuable evidence that aids in understanding a user's online activities.

\paragraph{Bookmarks analysis} Bookmarks in a web browser are links to websites and URLs that users save for easier access in the future. These bookmarks can reveal the sites a user frequently visits or intends to revisit later. By analyzing bookmarks, investigators can gain insights into a suspect's motivations and the types of websites they are interested in \citep{26}. This analysis can provide valuable information regarding the suspect's online behavior and preferences. Overall, bookmark analysis is a crucial technique in digital forensics, as it helps uncover digital evidence related to web browsing activities and user habits.

\paragraph{Download analysis} Download analysis involves examining the downloaded files by the suspected user on the device, which include documents, images, audio, videos, and other related files from the web. This technique helps in identifying the motivations behind the downloads and provides valuable insight into the types of files obtained. Such files can reveal the suspect’s interests and activities, potentially leading to incriminating evidence relevant to the investigation \citep{3}.

To perform download analysis, specialized digital forensics software can be used to extract and organize the downloaded files from the device, analyze patterns, and categorize the types of files downloaded. Overall, download analysis is an important technique in digital forensics as it helps investigators to uncover digital evidence related to file downloads and user behaviors.

\paragraph{Password analysis} Password analysis involves extracting information related to the credentials of suspected users, including usernames that browsers store for convenience during site visits. These usernames often auto-fill when the user returns to a site, while passwords can be saved for easier access during logins. This technique offers valuable insights into the user's online activities and may reveal the types of information and data the user stores and accesses . Additionally, password analysis can expose vulnerabilities, aiding in the potential cracking of passwords for multiple accounts. Specialized password analyzer software can be utilized to extract and scan passwords stored in browsers.

\paragraph{Malware analysis} This technique involves scanning and examining the suspected user's device for malicious code and executable files. A thorough investigation into the source and nature of the malware can provide insights into the origin of the malicious software and the websites from which it was downloaded. Malware forensics software can identify potentially vulnerable code within suspicious scripts, enabling investigators to determine the file's purpose \citep{3,31}. Conducting this analysis is crucial for understanding how cybercrimes and data breaches occur, offering insights into the attacker's tactics, techniques, and procedures, as well as potential vulnerabilities within the system or network. 

\paragraph{Live analysis} This technique is essential in digital forensics as it allows investigators to conduct real-time analysis on the suspect's device without disrupting ongoing applications or closing the device. During this phase, specialized applications monitor user activity and collect data from active programs, providing immediate insights into the suspect's actions. Live analysis is crucial because it evaluates the overall system, including all running applications, while also examining system logs and data residing in the device's current memory. This approach ensures that the suspect cannot delete previous data or alter device settings, allowing investigators to obtain firsthand evidence that can be vital for the case \citep{30}.
    
\paragraph{Memory analysis} The memory analysis forensics technique focuses on examining the random-access memory (RAM), which serves as the device's primary storage for volatile data associated with browser applications and other processes~\citep{26}. During this phase, investigators utilize specialized software to scan and analyze the memory, extracting data that may include information about open applications, accessed files, and other critical details related to browser usage. This technique can uncover hidden or encrypted files residing in memory. Memory analysis is a vital aspect of digital forensics, as it enables investigators to retrieve valuable information that may not be accessible through traditional disk-based forensics, proving essential in cases involving complex or sophisticated attacks \citep{32}.

\paragraph{Cloud-based and extension analysis} The cloud-based technique investigates information captured by applications, particularly online cloud storage, where users upload, share, and download files. User activities are recorded, revealing interactions made by the suspected individual, including details about accessing, sharing, and downloading files, as well as timestamps and file sizes. Additionally, browser plugins and extensions provide insights into the types of websites accessed by the user; each extension serves a specific purpose, such as hiding the IP address, blocking ads, or safeguarding the user's identity while browsing. This information assists investigators in identifying the user’s activities and understanding their behavior. Cloud-based and extension analysis are crucial techniques in digital forensics, as they enable the extraction of valuable data from cloud services and the examination of web browser extensions, which can be critical in investigations related to cybercrime or online fraud \citep{30}.

\section{Related work}\label{sec:related-work}
To gain a deeper understanding of existing research and identify potential gaps in knowledge, this section reviews relevant scholarly literature on various studies that have examined the effectiveness of browsing modes in preventing data retention and tracking across popular browsers. 

One of the foundational studies by \citep{4} scrutinized the private browsing modes of major browsers, including Google Chrome, Mozilla Firefox, and Microsoft Edge. Despite the design intent of these modes to prevent data retention, the research found that residual data could still be recovered using forensic techniques. This study employed tools like FTK Imager and EnCase, uncovering persistent artifacts such as cache files, cookies, and user activity logs. While it emphasized the limitations of forensic tools in accurately retrieving data from private browsing sessions, it also highlighted a critical gap in the privacy features available on Android platforms. However, this research primarily focused on desktop environments, potentially overlooking the nuances of mobile browsing behaviors.

In contrast, another study \citep{5} expanded the evaluation to 14 desktop browsers, including Brave and Tor, across both Windows and macOS platforms. This study categorized browsers into high, moderate, and low privacy protection tiers based on their default configurations. It was found that Brave, LibreWolf, and Tor offered the highest level of privacy, with Brave showing notable resistance to fingerprinting. Tor provided strong anonymity but at the cost of slower performance due to onion routing. DuckDuckGo, Firefox, Waterfox, and Safari were rated as moderate, with some vulnerabilities in tracking and content blocking. Conversely, Chrome, Edge, and Opera were deemed to offer minimal privacy safeguards. The study also noted that MacOS generally provided better privacy protection than Windows and recommended robust default privacy settings and open-source browsers as viable options for novice users.

\citet{6} evaluated the effectiveness of private browsing modes in Mozilla Firefox and Google Chrome across various Linux environments, utilizing forensic tools such as Autopsy for file system analysis, FTK Imager for creating forensic images, browser-specific forensic tools for analyzing browser artifacts, and manual analysis techniques along with custom scripts for extracting and analyzing data. The research discovered that while private browsing modes prevent disk storage of browsing information, significant data such as search keywords and login credentials can still be recovered from memory, especially in virtualized environments, even after a system restart. Non-virtualized environments exhibited fewer recoverable artifacts. The study also noted that Firefox allows new entries in private mode, whereas Chrome does not, and emphasized the importance of considering memory and swap space in privacy assessments. The findings indicate that private browsing modes are not entirely secure and can expose sensitive information through memory artifacts.

Another study \citep{3} suggested a comprehensive examination was conducted on Mozilla Firefox, Google Chrome, and TOR Browser. The main forensics tool used was FTK and the environment as in a virtual machine that had the Windows 7 operating system. Most of the data gathered when using the browsers are from normal mode. However, the data that was recovered in the private mode was comparatively less. Furthermore, almost no data can be retrieved when using TOR Browser. It should be noted that this analysis exclusively focused on the recoverable data from a hard disk image.

In \citep{33}, a forensic analysis of Mozilla Firefox, Google Chrome, and Microsoft Edge on Windows 11 focuses on artifact recovery across the entire browser lifecycle, including installation, browsing, crashes, and uninstallation. Key contributions include the development of a comprehensive methodology to identify artifacts from the registry, memory, logs, and storage during browser activities. The study found that Google Chrome retains the most sensitive data in memory, such as passwords and emails, while Microsoft Edge produces extensive registry and log artifacts. Firefox performs best in terms of privacy by leaving fewer artifacts. However, the analysis only examines the normal mode and does not test the other two modes private and portable. The paper concludes that despite improvements in browser security, significant forensic artifacts remain, underscoring the need for more advanced forensic tools to capture both volatile and persistent data effectively.

In the contact of social media, \citet{1} examined artifacts from the Discord application within Google Chrome. With the rise in popularity of social media platforms like Discord, this study examined various digital remnants, such as payment information, messages, and account settings, recovered from local storage and cache files. The findings revealed that significant forensic artifacts, including user IDs, login tokens, and chat logs, could be obtained from these sources. The research suggested that similar data storage mechanisms across desktop and web versions of Discord allowed for effective artifact recovery in virtual environments and recommended analyzing artifacts from Discord’s mobile applications for a more comprehensive understanding.

According to \citep{7}, a series of scenarios were designed and executed to evaluate the level of user privacy that was provided by the TOR Browser. Windows 8.1 operating system was used in the virtual machine environment where a three-stage process was followed for data acquisition. The web browser was launched, secondly, after some web activities like surfing the internet, and visiting sites were done then lastly the web browser was closed then the machine hard drive, RAM, and registry were thoroughly analyzed which revealed that the TOR Browser had left a considerable number of artifacts in the computer's memory which can be used to pull out the user's web activities. Moreover, \citep{8} evaluates the effectiveness of Google Chrome's Incognito mode by analyzing its local system interactions during private browsing sessions. Despite the common belief that Incognito mode ensures complete privacy, the investigation reveals that Chrome's Incognito mode does generate temporary files and performs write operations on the local disk, such as creating and deleting .tmp files. These findings are compared to standard browsing sessions, showing a significant reduction in local disk writes when using Incognito mode. However, the study also highlights potential privacy risks, as data written to these temporary files could potentially be recovered if not properly deleted. The paper further compares Chrome's Incognito mode with private browsing features in other browsers, emphasizing that while Incognito mode minimizes local data storage, it does not offer complete invisibility on the internet.

The study \citep{9} examined the browser artifacts present on a client computer within the same network as the TOR Browser, focusing on how these artifacts can provide insights into user activity and potential security implications. This research proposed a new technique for finding more valuable browser artifacts in other browsers such as Chrome Incognito, and Internet Explorer. Moreover, this study concludes that even the most renowned TOR Browser, which is known for its security and privacy, has left behind some traces.

A survey \citep{10} highlighted the growing need for effective forensic methods due to increasing cybercrimes. This review of current forensic tools and techniques identified challenges such as data structuring, result repeatability, and tool advancement but did not provide specific recommendations for addressing these issues in the context of different browsing modes.

\begin{table}[ht!]
\centering
\caption{Comparison of state-of-the-art studies}
\resizebox{1\textwidth}{!}{
\begin{tabular}{p{1cm}p{4cm}p{2.5cm}p{2cm}p{5cm}p{4.5cm}}
\hline
Papers & Browsers & Browser Modes & Test Scenario & Tools & Artifacts Recovered \\
\hline\hline
\citep{4} & Google Chrome, Brave, Mozilla Firefox, Tor Browser & Private & One & Browser History Capture, ADB, SQLite Database Browser, Hex Editor, Autopsy, Wireshark & Cookies, History, Cache, Session Data \\
\hline
\citep{33} & Mozilla Firefox, Google Chrome, Microsoft Edge & Normal, Crash Mode, Reset Mode & One & Regshot, MagnetRAMCapture, WinMerg, HashCalc, FTK Imager, HxD Hex Editor, Windows Logs Explorer & Registry Artifacts, Memory Artifacts, Storage Artifacts, Log Artifacts \\
\hline
\citep{5} & Brave, Chrome, Chromium, DuckDuckGo, Edge, Epic, Firefox, Libre Wolf, Opera, Safari, Tor Browser, Vivaldi, Waterfox, Yandex & Normal & Two & PrivacyTests online tools and BrowserLeaks & Browsers were tested in default configurations and results show for user privacy – low to high. \\
\hline
\citep{6} & Google Chrome, Mozilla Firefox & Private & Four & Forensic Toolkits, Browser-Specific Tools, Manual Analysis Techniques, Custom Scripts & Session Data, Cache, History, Metadata, Logs \\
\hline
\citep{3} & Tor Browser, Google Chrome, Mozilla Firefox & Private & One & Forensic Analysis, NTUSER.dat Hive, Live Systems Analysis & User Assist Keys, State File, Torcc File, Extensions, Packaged Software Versions, PAGEFILE.SYS Artifacts \\
\hline
\citep{1} & Google Chrome & Normal & One & Virtual VMs, Disk Analysis, Recovery Tools & Cache, Logs, Chats, Emojis, and other Discord-related data \\
\hline
\citep{7} & Tor Browser & Normal & One & MiniTool Partition Wizard, Regshot, Volatility, Hex Workshop, AccessData FTK Imager v4.1.1.1, Magnet AXIOM, Bulk Extractor & Web Cache, Cookies, Temporary Files, Browser History, Local Storage \\
\hline
This study & Brave, Tor, Mozilla Firefox, Google Chrome, Microsoft Edge & Normal, Private, Portable & Four & AccessData FTK Imager, SQLite, WinHex, BrowsingHistoryView, DB.BrowserSQLite, ChromeCacheView, MZCookiesView, Web Historian, Volatility, DCode & Places.sqlite, Cookies.sqlite, Formhistory.sqlite, Logins.json, Thumbnails (Images), History, Bookmarks, Downloads, Search Keywords, Top Sites, Visited Links, Login Data \\
\hline
\end{tabular}
}
\label{tab:1}
\end{table}

\newpage
Table \ref{tab:1} provides a comparative overview of recent studies in browser forensics, highlighting the different approaches researchers take in analyzing various web browsers and browsing modes. The majority of these studies have focused on either normal or private browsing modes, using diverse forensic tools to recover artifacts such as cookies, history, cache, and session data. Some only examine one or two browser modes, as seen in studies like \citep{4,6,3}, which focus on private mode exclusively or only cover specific modes like normal and crash/reset mode as in \citep{33}. However, several studies have notable limitations in scope, as they either do not explore portable browsing modes or fail to comprehensively cover all test scenarios involving data retention and recovery.

Our study fills these gaps by extending the analysis to all three browser modes—normal, private, and portable—using four distinct test scenarios. This paper uniquely addresses forensic data acquisition across these modes, examining artifacts such as browsing history, cookies, form data, login details, and cache using advanced forensic tools. Our approach not only provides a more holistic view of browser behavior across different modes but also aims to enhance the accuracy of forensic investigations by focusing on a wider range of browsing configurations and potential data leakage points. This highlights the need for further research to address the gaps, improve privacy measures, and develop more robust forensic methodologies to enhance overall data protection and forensic accuracy.

In this research, we examine both popular and emerging web browsers that prioritize user privacy and security across various modes, such as normal, private, and portable. Despite periodic updates introducing new privacy and security features, many artifacts can still be recovered, revealing potential vulnerabilities in each mode. Our objective is to analyze different web browsers using various computer forensic tools to recover the types of browser artifacts left behind on devices in these different modes. To guide this investigation, we address the following research questions:

\begin{enumerate}[label=\textbf{R\arabic*.}, leftmargin=*, labelwidth=*, align=left]
    \item What artifacts and data are recovered from each distinct browser after utilizing various forensics tools and techniques, and to what extent can these artifacts be used as evidence?
    \item Which tools are best suited for conducting web forensics acquisition?
    \item What is the most effective mode for browser data acquisition for each browser?
\end{enumerate}

\section{Methodology}\label{sec:methodology}

\begin{figure}[!htb]
    \centering
    \includegraphics[width=1\linewidth]{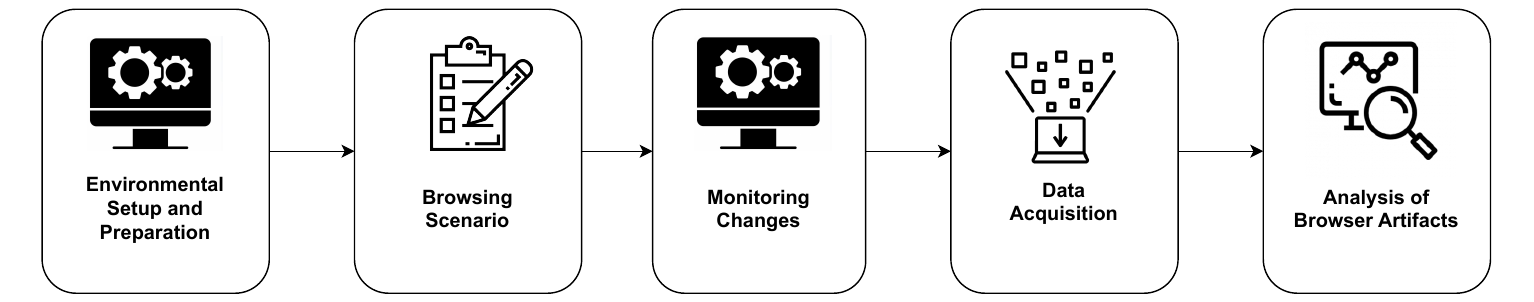}
    \caption{Forensics methodology}
    \label{fig:2}
\end{figure}

The methodology presented in this paper addresses the research questions aimed at evaluating the effectiveness of web browsers using digital forensics tools to identify artifacts recoverable in normal, portable, and private browsing modes. Research questions R1, R2, and R3 are addressed through this methodology. It comprises five key phases, as illustrated in Figure \ref{fig:2}: (i) environment setup, (ii) use case scenario, (iii) monitoring changes, (iv) data acquisition, and (v) analysis of browser artifacts.



\subsection{Environmental setup and preparation}
The objective of this phase was to deploy a clean virtual environment where the browsers and forensics tools were downloaded and installed. Each browser was tested in this controlled environment, ensuring it was used for the first time, meaning no prior user data or cache was available. In each scenario, the web browser was set as the default and used for normal private, and protable browsing activities, including general searches, downloading images, videos, documents, and user logins. 

The study presented in this paper was conducted on an Acer Aspire 3 laptop running the Windows 11 operating system. Oracle Virtual Box machine (v7.0.8-156879-Win) was used to create a virtual environment created on a Windows 11 (64-bit) operating system, equipped with 6 GB of RAM and 20 GB of storage. All updates were completed, and Windows updates were disabled to ensure the virtual environment remained stable. Following the environment setup, the next step was the browser selection process. Table \ref{tab:3} presents a list of various browsers selected for this research based on their popularity and security features. Each browser was installed in the virtual environment with default configurations and tested on the normal, private, and portable browsing modes. 
\begin{table}[!ht]
\centering
\caption{List of browsers used in the experiment }
\begin{tabular}{ll}
\hline
Browser & Version \\
\hline\hline
Brave & 1.51.114 \\

TOR Browser & 12.0.4 \\

Mozilla Firefox & 113.0 \\

Google Chrome & 113.0.5672.92 \\

Microsoft Edge & 113.0. 1774.35 \\
\hline

\hline
\end{tabular}
\label{tab:3}
\end{table}

Each browser stores user data and information based on the browsing session, and each type of recoverable artifact is located in specific directories on the computer. In this methodology, it is essential to re-examine and analyze various data points to identify and gather insights into user intentions and uncover potential evidence in the form of browser artifacts. Table~\ref{tab:4} provides an overview of the types of artifacts considered, while Table~\ref{tab:5} lists the forensic tools used in the experiment. 

\begin{table}[h!]
\centering
\caption{Potentially recoverable contents}
\begin{tabular}{ll}
\hline
Recoverable Contents & Location \\
\hline\hline
Websites Accessed & Browser History, Cookies, Temporary files/folders, Cache \\

Timestamps & Cookies, Cache, Browser History \\

Search Queries/Keywords & Browser Auto-fill Data, Suggestion, Auto Compete, Cache  \\

Websites Saved & Browser Bookmarks/Favourites \\

Downloads & Downloads Section, Cache \\

Extensions & Browser Extensions Manager \\
\hline

\hline
\end{tabular}
\label{tab:4}
\end{table}

\begin{table}[!ht]
\centering
\caption{Tools used in the experiment}
\begin{tabular}{lcp{10cm}}
\hline
Tools & Version & Purpose\\
\hline\hline
AccessData FTK Imager & 4.7.1 & Creation of forensic disk and memory images/ Browses artifacts, directory folders, and pagefile.sys, hiberfil.sys \\

SQLite & 3.42.0 & Browse disk space databases, browser SQL files \\

WinHex  & 20.8 & Hexadecimal data recovery  \\

BrowesingHistoryView & 2.55 & Browses artifacts, directory folders \\

DB. BrowserSQLite & 3.12.2 & Open/View browser SQL files.\\

ChromeCacheView & 2.45 & View Google Chrome cache \\

MZCookiesView & 1.60 & Cookies analyzer \\

Web Historian & 2.0 & Analyze history\\

Volatility & 2.6 & Memory dump analysis \\

DCode & 5.5.21194.40 & Decode timestamps \\
\hline

\hline
\end{tabular}
\label{tab:5}
\end{table}

\subsection{Browsing scenarios}

\begin{figure}[!ht]
    \centering
    \includegraphics[width=1.0\linewidth]{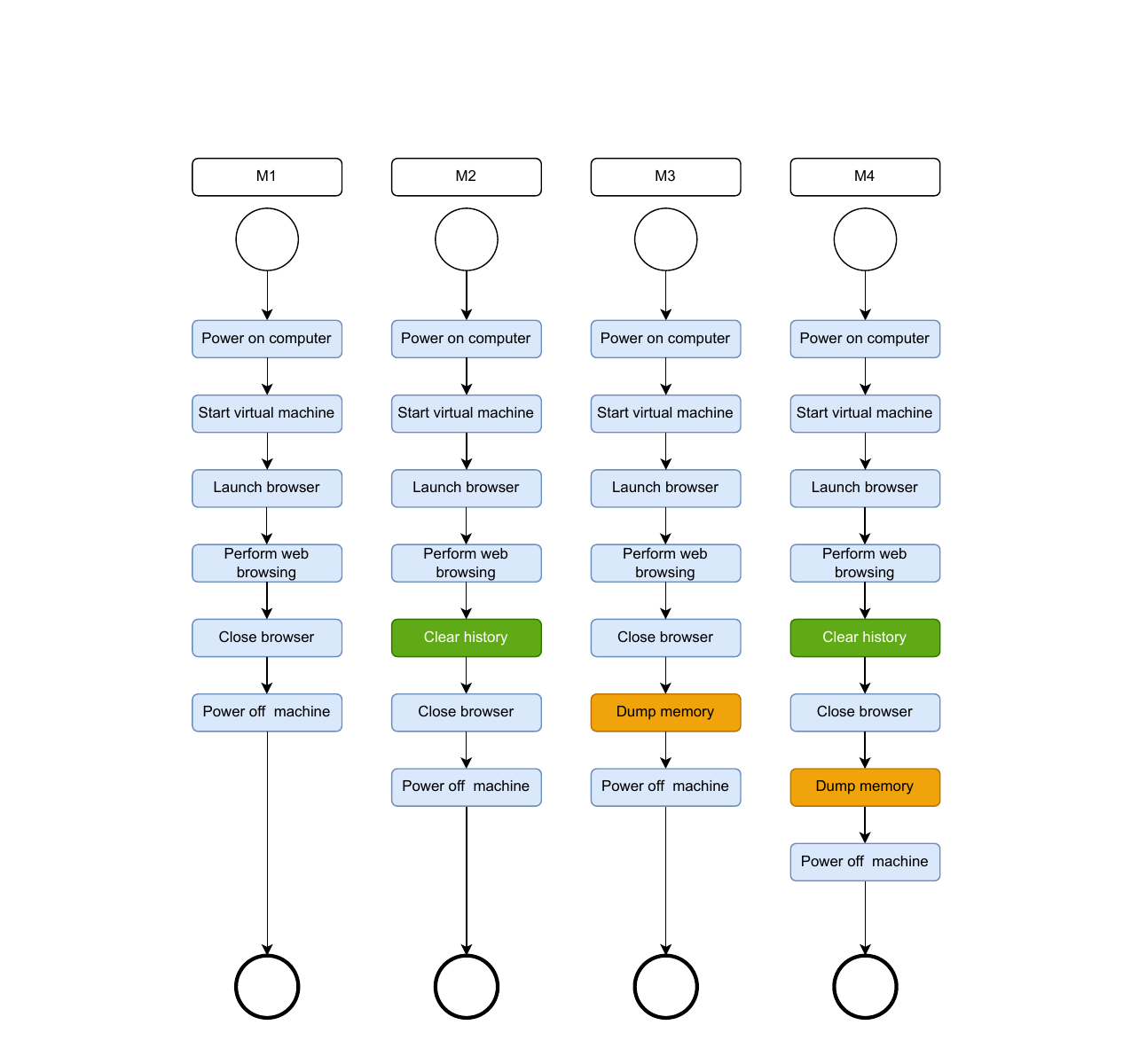}
    \caption{Browsing scenarios}
    \label{fig:3}
\end{figure}

Each browsing session involves distinct scenarios in which the user accesses the machine and performs various browsing activities in normal, private, and portable modes. We have designed four browsing scenarios, M1, M2, M3, and M4, to simulate these browsing sessions, as illustrated in Figure~\ref{fig:3}. 
For instance, browsing scenario M1 consists of six stages: turning on the computer, starting the virtual machine, opening the browser, engaging in web browsing and associated tasks, closing the browser, and shutting down the machine. Scenario M2 builds on M1 by adding a `clear browsing history' step between web browsing and closing the browser. Scenario M3 adds a `dump memory contents' step between closing the browser and shutting down the machine. Finally, scenario M4 combines both the `clear history' and `dump memory' stages from M2 and M3.

These scenarios were executed for each browser in all browsing modes, enabling a comprehensive analysis of the recoverable data at each stage. By monitoring each browsing session in a controlled environment, we can systematically evaluate the browser's state and data usage, ensuring consistent forensic analysis and extraction.
There are numerous web browsing activities that can be performed. As stated in \citep{8}, the most common browsing activities include web surfing, downloading files, viewing content in read mode and streaming multimedia. Additionally, \citep{6} discusses emerging web browsing activities, such as logging into websites, and entering URLs in the search bar without accessing the site. Table \ref{tab:6} provides a detailed list of the browsing activities used in our experiments.


\begin{table}[!ht]
\centering
\caption{Browsing keywords and activities}
\begin{tabular}{llp{6cm}}
\hline
Website & Keywords & Activities \\
\hline\hline
https://www.google.com/ & Cats & Searching for ``Cats" \\
\hline
https://www.bing.com/ & Cats & Searching for ``Cats" \\
\hline
https://images.google.com/ & Cat images & Searching for ``Cat images" \\
\hline
https://www.bing.com/images/ & Cat images & Searching for ``Cat images" \\
\hline
https://www.google.com/ & AE110 outer door handles & Searching for ``AE110 outer door handles" \\
\hline
https://www.bing.com/ & AE110 outer door handles & Searching for ``AE110 outer door handles" \\
\hline
https://www.google.com/ & Fiji travel destination places & Searching for ``Fiji travel destination places" \\
\hline
https://www.bing.com/ & Fiji travel destination places & Searching for ``Fiji travel destination places" \\
\hline
https://unsplash.com/ & Cats & Searching, viewing results, and downloading images (JPEG \& PNG) \\
\hline
https://www.youtube.com/ & UniFiji SOST & Search and stream ``UniFiji SOST videos" and stream various videos \\
\hline
https://www.facebook.com/ & Login Page & Logging in with user credentials, browsing through posts, groups interaction, and likes/comments \\
\hline
https://www.instagram.com/ & Login Page & Logging in with user credentials, browsing through posts, groups interaction, and likes/comments \\
\hline
https://mail.google.com/mail/u/0/ & Login Page & Logging in with user credentials \\
\hline
https://en.savefrom.net/387/ & Online video downloader & Download videos from YouTube \\
\hline
https://www.freepik.com/ & Party designs & Searching \& downloading assets with different file formats with zipped folders \\
\hline
https://www.unifiji.ac.fj/sost/ & Fees & Browsing and downloading pdf and document files \\
\hline
https://www.vitikart.com.fj/ & iPhone 14 & Searching for ``iPhone 14" and browsing through the other items \\
\hline
https://www.jbhifi.com.au/ & JBL Speaker & Searching for ``JBL Speaker" and browsing through the other items \\
\hline

\hline
\end{tabular}
\label{tab:6}
\end{table}

\subsection{Monitoring changes}
The monitoring phase focuses on investigating each stage of forensic analysis across different browsing sessions for each browser in normal, private, and portable modes. Each browsing session was conducted within a controlled virtual environment, followed by ananalysis of the browser's artifacts after the virtual machine was powered off. Additionally, we monitored the behavior of various forensic tools during data analysis and extraction to observe how each tool performed while retrieving browser artifacts.

\subsection{Data acquisition}
The goal of this phase is to collect data from the virtual machine and the various forensic tools used for data extraction. This data is typically obtained in raw or tool-specific formats, often representing the initial acquisition stage. After the data is collected, it is further analyzed and interpreted. The data primarily consists of browser artifacts, although some tools may retrieve incomplete or irrelevant information.

For browser data acquisition, we employed forensic browser analysis tools, SQLite database analysis, and browser-specific extraction techniques. These tools are specifically designed to analyze and gather browser-related artifacts. Since many browsers rely on a common SQLite database, critical data can be retrieved effectively. Data is collected from the primary locations where browsers store information on the computer's drive, as well as through tool execution, ensuring a comprehensive approach to artifact recovery.

\subsection{Data analysis} 

In this phase, we examine and interpret the data and information collected by various forensic software tools, focusing on browser artifacts retrieved from different browsers. The objective is to carefully extract sensitive and valuable data that can provide insights into the extent to which each browser retains information about a user's browsing session. This analysis encompasses different types of web-related artifacts generated during active browsing sessions.
The key areas of analysis include:
\begin{itemize}
\item \textit{Browser Artifact Analysis}: Identifying and evaluating specific artifacts stored by the browser, such as history, cache, cookies, and downloads.
\item \textit{Timeline Analysis}: Reconstructing the sequence of events during a browsing session to establish a timeline of user activities.
\item \textit{Content and Keyword Analysis}: Reviewing the content and search terms used by the user to understand their browsing behavior and interests.
\end{itemize}

By analyzing all the collected data, we aim to reconstruct the user's browsing activities and identify potential evidence. The collected artifacts were then compared to evaluate the performance of different forensic tools across various browsers and browsing modes.

\section{Analysis of Browser Artifacts}\label{sec:analysis}
Browser artifact data were collected in three browsing modes--normal, private, and portable--after executing the four browsing scenarios (M1, M2, M3, and M4).

\subsection{Normal Browsing Mode }
\begin{figure}[!ht]
    \centering
    \includegraphics[width=0.9\linewidth]{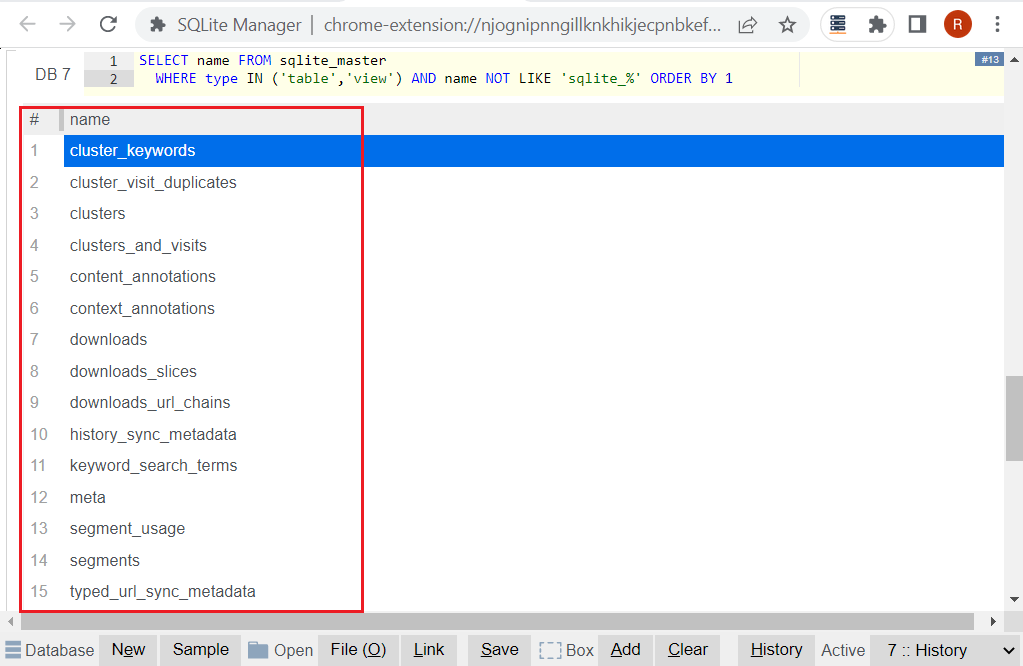}
    \caption{Screenshot of SQLite for Mozilla Firefox}
    \label{fig:4}
\end{figure}

In the M1 scenario, various forensics tools were used to analyze and extract browsing data and artifacts. The SQLite database revealed information captured from the overall browsing activity, which was stored in the root folder path: \texttt{...\textbackslash AppData\textbackslash Local\textbackslash Google\textbackslash Chrome\textbackslash User Data\textbackslash Default}. Within this directory, the \texttt{History} file, as shown in Figure \ref{fig:4}, contains recoverable data on web history, including keywords and site visits conducted during the browsing session.

\begin{figure}[!htb]
    \centering
    \includegraphics[width=0.88\linewidth]{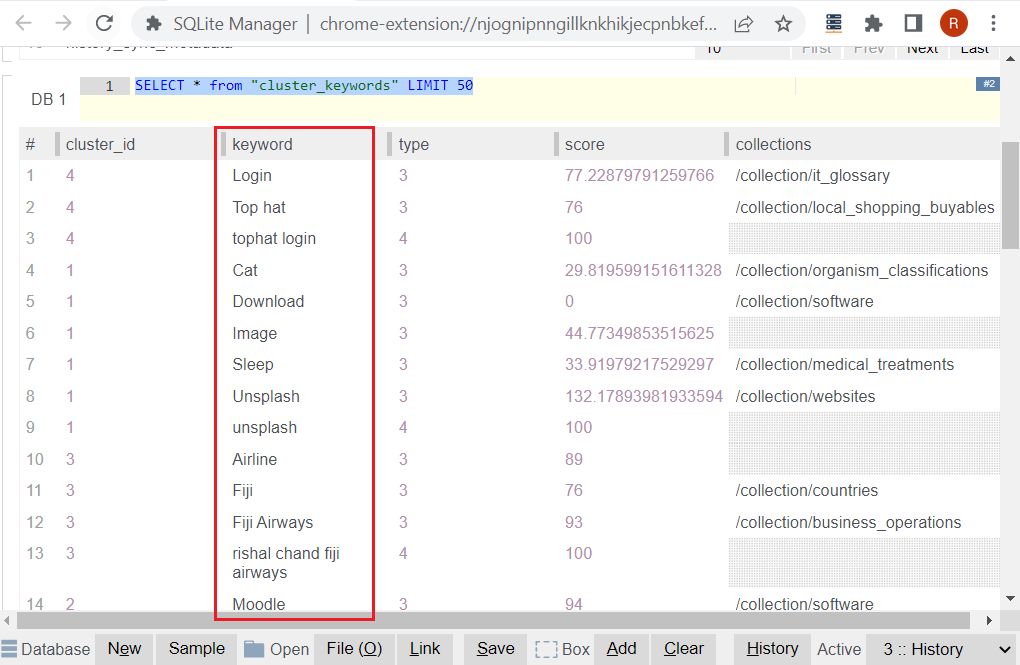}
    \caption{Screenshot of retrieved SQLite keywords of Mozilla Firefox}
    \label{fig:5}
\end{figure}

A total of 50 cluster keywords were selected from the \texttt{History} file during testing in normal browsing mode, as shown in Figure \ref{fig:5}. The data includes fields such as \texttt{cluster\_id}, \texttt{keyword}, \texttt{type}, \texttt{score}, and \texttt{collections}. Since the search results are not arranged in a specific order, the file reflects the score and category collections for each keyword used during the browsing session.

\begin{figure}[!htb]
    \centering
    \includegraphics[width=0.88\linewidth]{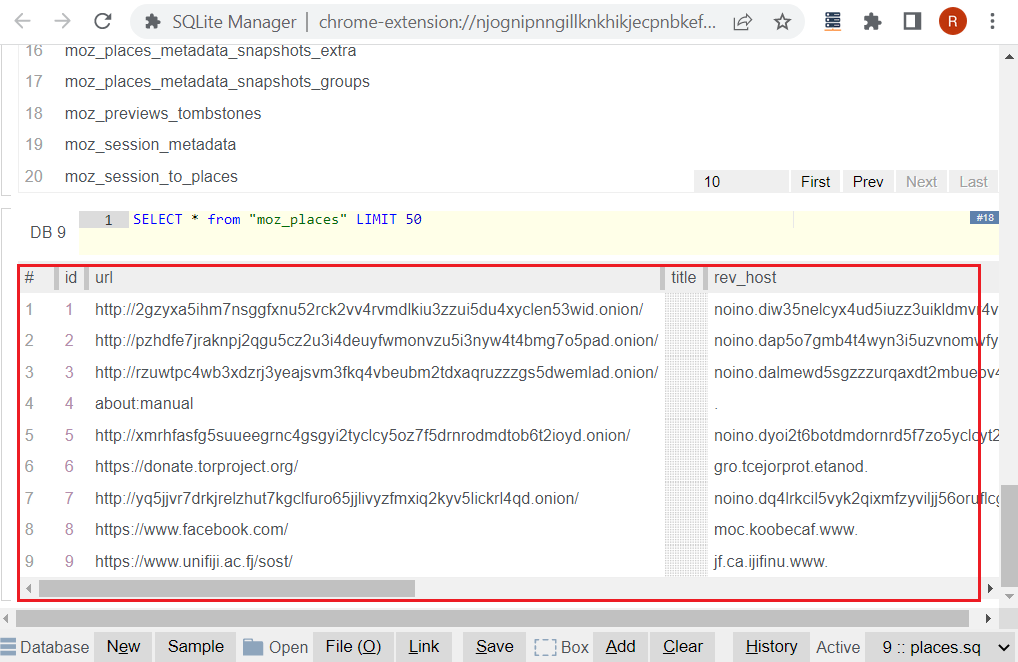}
    \caption{Screenshot of SQLite manager of TOR browser data}
    \label{fig:6}
\end{figure}

The same browsing activities were conducted on other browsers, including Brave, Mozilla Firefox, and Microsoft Edge. These browsers similarly saved session history, containing details such as the \texttt{URL}, \texttt{title}, \texttt{visit time, visit count, referring page, visit type, visit duration, browser type, user profile, browser profile, URL length, typed count, history file}, and \texttt{record ID}, as seen in Figure \ref{fig:6}. Since this was conducted in M1, the browser was closed after each session. Forensic tools such as SQLite, ChromeCacheView, BrowsingHistoryView, and MzCacheView were used, yielding comparable results across all tested browsers.

During the testing of M2, M3, and M4, the history and cache of each browser were cleared, and storage locations were manually deleted to remove any remaining data. Despite these efforts, fewer artifacts were recovered at each stage. However, even in normal browsing mode, some artifacts remained recoverable. These include keywords, URLs, session thumbnails, and cache files. 

\subsection{Private Browsing Mode}
\begin{figure}[!htb]
    \centering
    \includegraphics[width=0.8\linewidth]{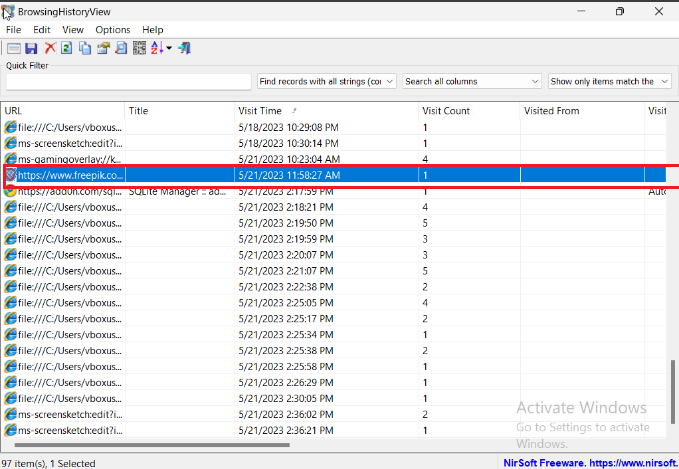}
    \caption{Screenshot of browsing history view -- private mode Brave browser data}
    \label{fig:7}
\end{figure}

In private browsing mode, all browsers claimed to prevent tracking and recording of user activities. However, while using tools like ChromeCacheView and MzCacheView no results were found. The exception was BrowsingHistoryView, which recorded a downloaded file from Brave browser, as shown in Figure \ref{fig:7}. 

\begin{figure}[!ht]
\centering
\begin{subfigure}{0.49\textwidth}
    \includegraphics[width=\textwidth]{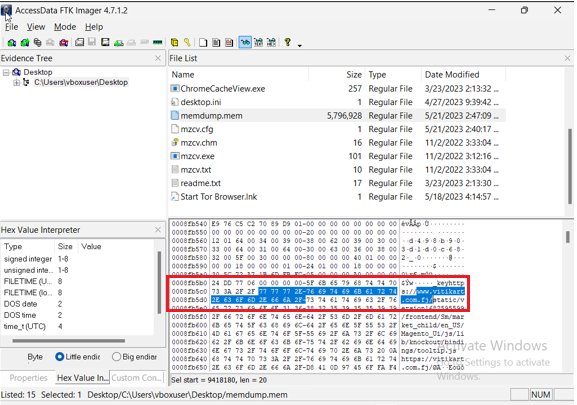}
    \caption{}
    \label{fig:ftk1}
\end{subfigure}
\hfill
\begin{subfigure}{0.49\textwidth}
    \includegraphics[width=\textwidth]{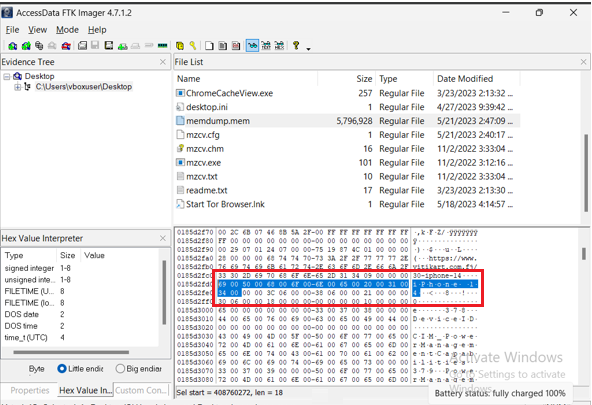}
    \caption{}
    \label{fig:ftk2}
\end{subfigure}
\caption{Screenshots of AccessData FTK Imager showing (a) memory dump retrieval and (b) analysis results}
\label{fig:8}
\end{figure}

Additionally, after conducting the M2, M3, and M4 scenarios, some artifacts were still recoverable. These included fragments of information such as keywords, URLs, and broken text strings from certain activities. Figure~\ref{fig:8} shows the keywords, URLs, and other data retrieved during the searches. 

\subsection{Portable Mode}
\begin{figure}[!htb]
    \centering
    \includegraphics[width=0.8\linewidth]{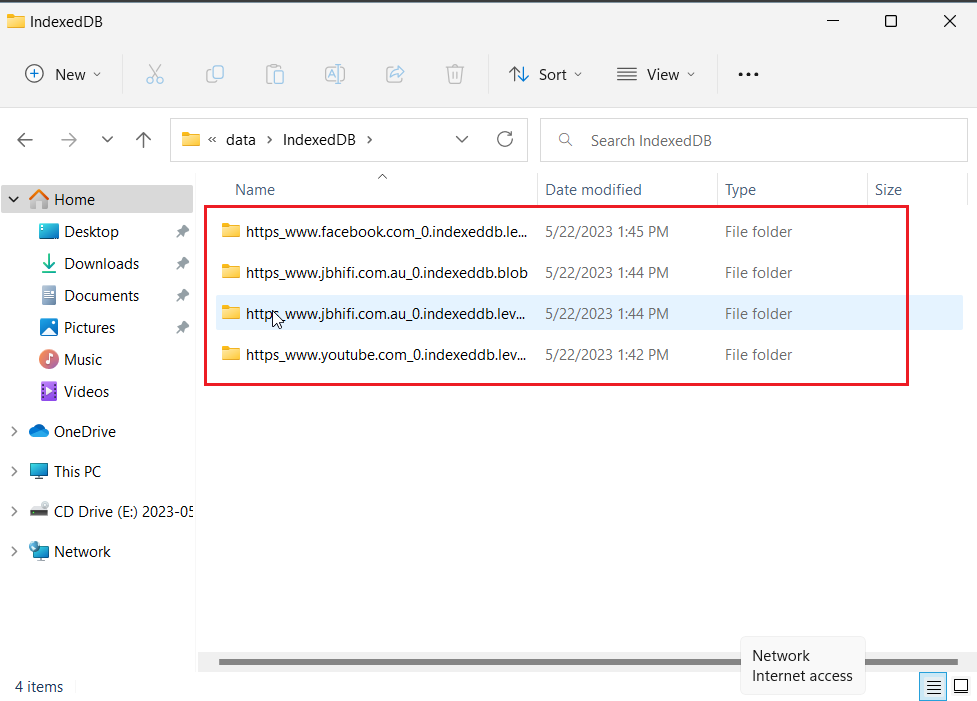}
    \caption{Screenshot of Brave Browser IndexedBD }
    \label{fig:9}
\end{figure}

For portable mode testing, previously installed browsers were uninstalled, and their folders were manually deleted. Portable browsers were then downloaded, extracted, and tested according to scenarios M1, M2, and M3. The results revealed that even portable browsers retained keywords, URLs, and other information from browsing sessions. \textit{Indexed DB} displayed the sites visited in the Brave browser, while \textit{AccessData FTK Imager} revealed the URLs stored in Local Storage, as shown in Figure \ref{fig:9}.

\begin{figure}[!htb]
    \centering
    \includegraphics[width=0.9\linewidth]{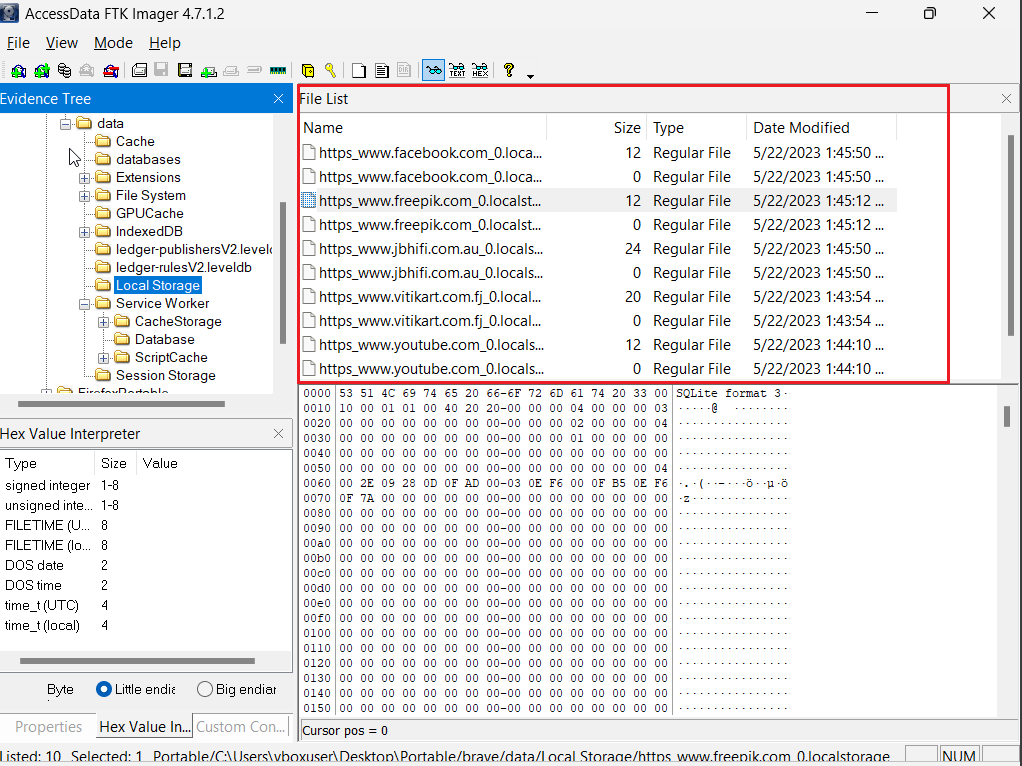}
    \caption{Screenshot of AccessData FTK Imager - Brave Browser local storage}
    \label{fig:10}
\end{figure}

\textit{AccessData FTK Imager} also displayed a file list of visited URLs and retrieved keywords for Brave Browser, as illustrated in Figure \ref{fig:10}.

\begin{figure}[!ht]
\centering
\begin{subfigure}{0.49\textwidth}
    \includegraphics[width=\textwidth]{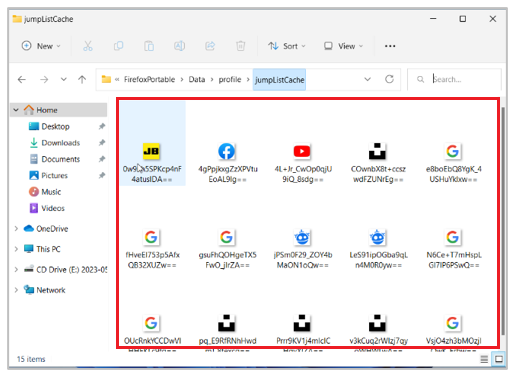}
    \caption{}
    \label{fig:firefox1}
\end{subfigure}
\hfill
\begin{subfigure}{0.49\textwidth}
    \includegraphics[width=\textwidth]{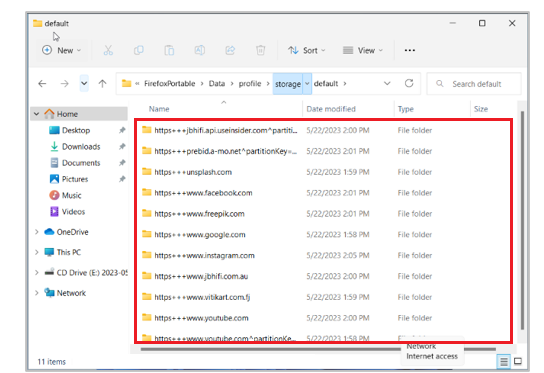}
    \caption{}
    \label{fig:firefox2}
\end{subfigure}
\caption{Screenshots of the portable folder location in Mozilla Firefox showing (a) thumbnails and (b) default storage location}
\label{fig:112}
\end{figure}

After performing M3, the portable browser's root directory retained thumbnails and website visit information. These files were accessible in the portable browser's folders, as shown in Figure \ref{fig:112}. 

\begin{figure}[!htb]
    \centering
    \includegraphics[width=0.9\linewidth]{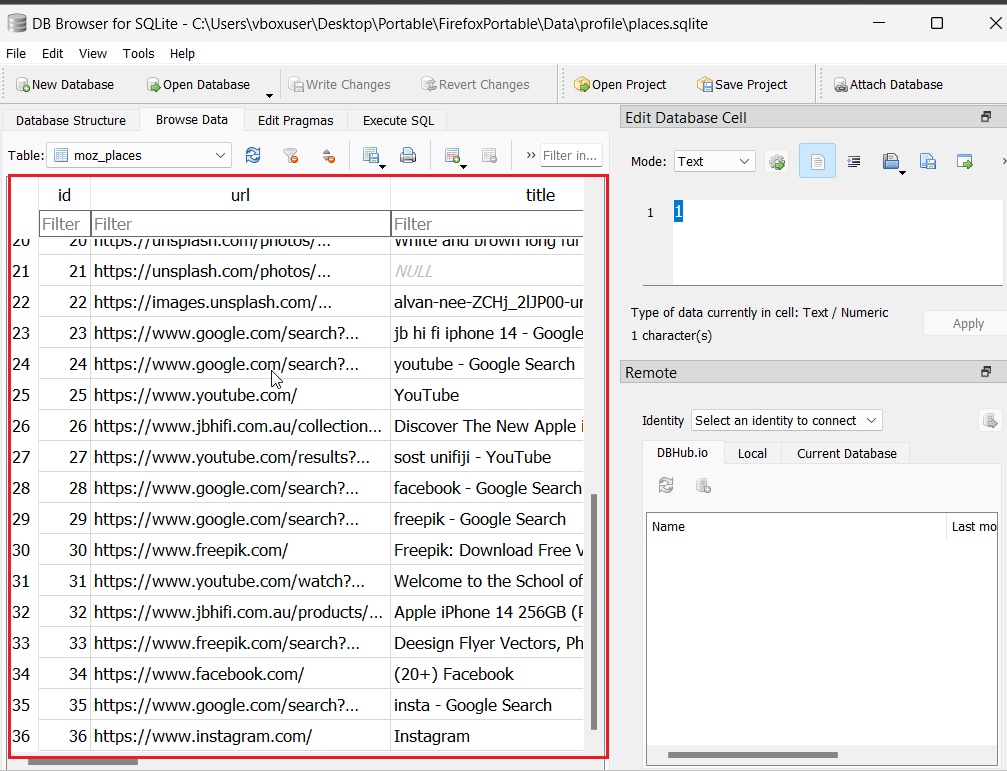}
    \caption{Screenshot of DB Browser for SQLite -- places visited}
    \label{fig:places}
\end{figure}

\begin{figure}[!htb]
    \centering
    \includegraphics[width=0.9\linewidth]{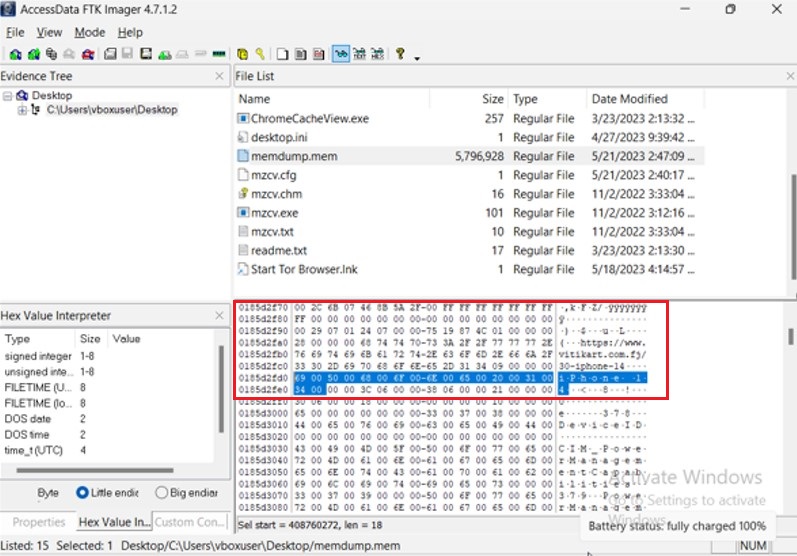}
    \caption{Screenshot of AccessData FTK Imager cookies view - Microsoft Edge}
    \label{fig:13}
\end{figure}

Despite manually clearing the cache and history, performing a memory dump, and restarting the machine, some temporary files remained in the portable root folder. These files included the downloaded files and a \texttt{places.sql} file that contained the URLs of the browsed sites, as illustrated in Figure~\ref{fig:places}. Cookies file retained information about the websites that were visited as well as shown in Figure~\ref{fig:13}.

\section{Evaluation Results}\label{sec:results}

\begin{figure}[!ht]
\centering
\begin{subfigure}{0.48\textwidth}
    \includegraphics[width=\textwidth]{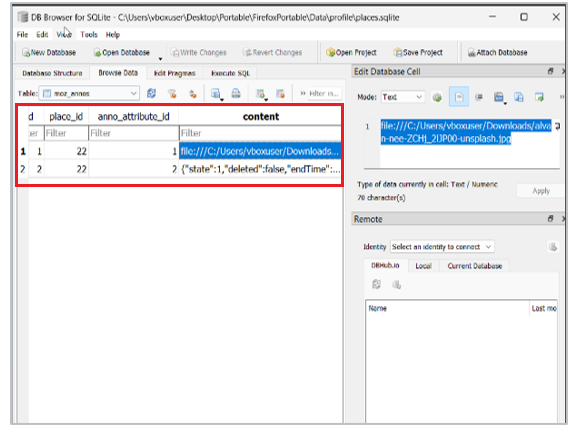}
    \caption{}
    \label{fig:db1}
\end{subfigure}
\hfill
\begin{subfigure}{0.48\textwidth}
    \includegraphics[width=\textwidth]{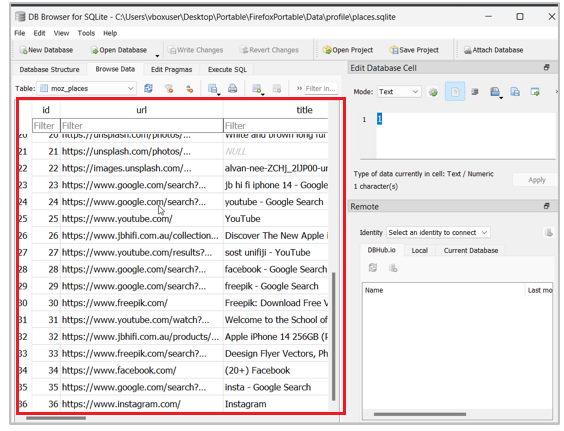}
    \caption{}
    \label{fig:db2}
\end{subfigure}
\caption{Screenshots of DB browser for SQLite showing (a) profiles and (b) places visited}
\label{fig:14}
\end{figure}

The evaluation revealed that multiple browsers retained significant information that could serve as digital forensics evidence. This data was locally saved within the virtual machine. As illustrated in Figure \ref{fig:14}, various browsers stored artifacts corresponding to each methodological scenario, with each browser leaving behind information related to web browsing sessions. This data was typically found in the default locations for each browser, whether used in normal or private browsing modes.

The information was saved in different file formats, including SQLite databases and JSON files. Most of the relevant files were accessed using SQLite DB Manager and the Browser Extension SQLite, with retrieval conducted through tools like WinHex and FTK Imager.

Mozilla Firefox stores various browser artifacts, including Browser History, Cookies, Bookmarks, Thumbnails, Add-ons, Downloads, and Add-on information, in the root location of the device using SQLite format. Additionally, some information is saved in JSON format. Each artifact is saved with a distinct filename within the \texttt{...\textbackslash Roaming\textbackslash Mozilla\textbackslash Firefox\textbackslash Profiles} folder, found in a randomly generated profile folder, and subsequently in the default folder. 

Thumbnails are located in the \texttt{...\textbackslash Local\textbackslash Mozilla\textbackslash Firefox\textbackslash Profiles} directory, specifically in the random profile folder, then in the default folder \textbackslash thumbnails. Bookmarks and cookies provide valuable insights into user activity, including search keywords that can help track and analyze suspected browsing history. Table \ref{tab:7} presents the results of the recovered browser artifacts from Mozilla Firefox, along with their corresponding filenames. Notably, most artifacts remain recoverable even after users clear their history and cache, as this information persists in the installation directory. Moreover, even after deleting files from the root directory, traces of browsing sessions can still be captured in the memory dumps.
\begin{table}[ht!]
\centering
\caption{Mozilla Firefox artifacts and forensic information}
\begin{tabular}{p{3cm}p{5cm}p{6cm}}
\hline
Filename & Mozilla Firefox Artifacts & Forensics Information\\
\hline\hline
Places.sqlite & browser history, bookmarks, and downloads & URLs, visit counts, Last visits, bookmark name, type, download paths, Timestamp, and Referrers. \\
\hline
Cookies.sqlite & Browser Session Cookies & Name, Path, Timestamp info, Site Session details, Keywords. \\
\hline
Extension & Addons, Plugins & Not available\\
\hline
Formhistory.sqlite & Browser Form History & Autofill, Names, and details. \\
\hline
Logins.json & Text file saved passwords/website login information & URLs, Usernames, Passwords, Emails (Encrypted) \\
\hline
Thumbnails (Images) & Website Thumbnails & Images related to websites visited. \\
\hline

\hline
\end{tabular}
\label{tab:7}
\end{table}

Similarly, the Google Chrome browser stores all artifacts, including history, downloads, preferences, visited sites, and cookies, in different files, most of which are formatted as SQLite files. Additionally, some files are saved as Session Saver (SNSS) format or as JSON files, located in the directory: \texttt{\textbackslash AppData\textbackslash Local\textbackslash Google\textbackslash Chrome\textbackslash User Data\textbackslash Default}. Bookmarks are also stored in a JSON format within the root directory, capturing details of websites by the user, thereby allowing for quick access in future sessions. The Downloads section logs all files downloaded, providing insight into the types of files the user has acquired. Moreover, keywords searched by the user while browsing different search engines are saved in the \texttt{History} file, offering additional context for tracking user activity. 

In Google Chrome's private browsing mode, there is typically less information available since the browser does not keep track of user activities or browsing history. However, some artifacts, such as search keywords, web cache, and visit sites, were still found in the memory dump on the device. This indicates that certain traces of user activity can be recovered. Additionally, Google Chrome logins can be accessed directly using Notepad, revealing usernames and providing a way to trace the sites visited based on these credentials. While some passwords are stored, they are encrypted and thus not readily accessible. Table~\ref{tab:8} summarizes the results for the artifacts recovered from the Google Chrome browser. 

\begin{table}[!ht]
\centering
\caption{Google Chrome artifacts}
\begin{tabular}{lll}
\hline
Filename & Google Chrome Browser Artifact & Forensics Information\\
\hline\hline
History & Browser History & URLs, Visit Counts, Last Visits \\

Bookmarks & User Bookmarks & Name, Path, Timestamp info, Site Timestamps \\

Cookies & Cookies & Session details, Keywords, Path \\

Downloads & User Downloads & Location path, File type, Referrer \\

Search Keywords & User Search Keywords & Sites visits, Searched keywords, Visits, Hits, URL \\

Top Sites & Top Sites & Top Sites visited and ranks \\

Visited Links & Visited Links & Links visited \\

Login Data & User Login Data & URLs, Usernames, Passwords, Emails (Encrypted) \\
\hline

\hline
\end{tabular}
\label{tab:8}
\end{table}

The Microsoft Edge browser stores browsing information in both SQLite and BAK file format, located in the directory:
\texttt{AppData\textbackslash Local\textbackslash Microsoft\textbackslash Edge\textbackslash User Data\textbackslash Default}. Within the default folder, several key files are stored, including History, Bookmarks, Favicons, Login Data, Media History, and Top Sites. The Downloads section provides valuable insights into the types of files downloaded and helps in identifying the websites from which these files were downloaded. Table \ref{tab:9} presents the recovered browser artifacts for Microsoft Edge.
\begin{table}[!ht]
\centering
\caption{Microsoft Edge browser artifacts}
\begin{tabular}{lll}
\hline
Filename & Microsoft Edge Browser Artifact & Forensics Information \\
\hline\hline
History & Browser History & URLs, Visit Counts, Last Visits \\

Bookmarks & User Bookmarks & Name, Path, Timestamp info, Site Timestamps \\

Cookies & Cookies & Session details, Keywords, Path \\

Downloads & User Downloads & Location path, File type, Referrer \\

Search Keywords & User Search Keywords & Site visits, Searched keywords, Visits, Hits, URL \\

Top Sites.sqlite & Top Sites & Top sites visited and ranks \\

Visited Links & Visited Links & Links visited \\

Login Data.sqlite & User Login Data & URLs, Usernames, Passwords, Emails (Encrypted) \\
\hline

\hline
\end{tabular}
\label{tab:9}
\end{table}

The Brave Browser also retains various types of browsing artifacts, which is stored in the directory: \texttt{AppData\textbackslash Local\textbackslash \\BraveSoftware\textbackslash Brave-Browser\textbackslash User Data\textbackslash Default}. Most artifacts, including Browser history, downloads, Bookmarks, Login Data, Network Action Predictor, and Shortcuts, are saved in SQLite and JSON formats. This information can be accessed using SQLite applications, which may be used in investigations against the suspected user. Downloads, keyword searches, and visited sites are key indicators that reveal the user's search history and can provide insights into their motivations. Table \ref{tab:10} lists the artifacts recovered from the Brave Browser.  

Most web browsers prioritize user privacy and security by offering features like private browsing mode, which prevents the tracking of history and overall browsing activity. Additionally, portable mode allows users to access browsers without leaving traces on the host machine. While private browsing is designed to ensure that no tracking occurs, it can also be exploited by individuals with malicious intent to obscure their online activities. Despite its promises of anonymity, suspects may take one step further to remove any remaining information, history, and downloads through the use of anti-forensics techniques. 
\begin{table}[!ht]
\centering
\caption{Brave browser artifacts}
\begin{tabular}{lll}
\hline
Filename & Brave Browser Artifact & Forensics Information\\
\hline\hline
History.sqlite & Browser History & URLs, Visit Counts, Last Visits \\

Bookmarks & User Bookmarks & Name, Path, Timestamp info, Site Timestamps \\

Cookies & Cookies & Session details, Keywords, Path \\

Downloads & User Downloads & Location path, File type, Referrer \\

Search Keywords & User Search Keywords & Terms, Searched keywords, Hits, ID \\

Top Sites.sqlite & Top Sites & Top Sites visited and ranks \\

Visited Links & Visited Links & Links visited \\

Login Data.sqlite & User Login Data & URLs, Usernames, Passwords, Emails (Encrypted) \\
\hline

\hline
\end{tabular}
\label{tab:10}
\end{table}

\section{Discussions}\label{sec:discussion}
Our experiments reveal that while browsers emphasize user privacy and security through their private browsing modes, they still retain some information on the device, even when users intentionally delete their browsing data. The application of various forensic tools successfully analyzed, identified, and gathered different artifacts from each browser, as summarized in Table~\ref{tab:artifacts}, across Normal, Private, and Portable modes of browsing.
The findings indicate that all browsers tested in this study leave behind some artifacts, suggesting that no browser is entirely safe from data recovery. This information is invaluable for investigators, as it provides insights into the types of recoverable artifacts that may assist in forensic analysis.
Below, we address the research questions posed at the beginning of this study:

\paragraph{\textbf{R1}} Multiple artifacts were recovered during the experimentation process, including URLs, search keywords, and other data, as detailed in Tables \ref{tab:7} and \ref{tab:8}. The results highlight that specific search keywords, particularly those with high-frequency hits, were identified using WinHex and FTK Imager tools from the memory dump. These tools also revealed additional artifacts such as URLs, top sites, and bookmarks.

While download history can be cleared, the path to the downloaded files remains accessible, allowing for the retrieval of files that can provide further insights. Additionally, web cache, cookies, and other residual files persist, contributing to the available data for analysis. The use of SQLite extensions and desktop applications like SQLite Manager has proven invaluable for exploring history, top sites, and login information.

Each type of artifact recovered from the browsers is significant, as it provides a glimpse into the suspect’s interactions with the browser. This information helps trace digital footprints, revealing the suspect's actions and understanding their web activity, which is crucial for establishing motives.


\paragraph{\textbf{R2}} The tools employed in this study are well-suited for performing web forensics acquisition, allowing for the analysis and identification of browser data locations and facilitating data acquisition. SQLite has proven to be particularly useful for accessing the SQLite database files that browsers use to store browsing data. One effective tool for this purpose is DB Browser, which enables users to view the database and execute queries. Beyond simple viewing, DB Browser supports full database administration activities typically performed by a database administrator; however, for web browser forensics, it primarily facilitates running queries to select the required tables.

FTK Imager is another essential utility that allows for the accurate duplication of a selected disk, whether physical or logical. It can create image files that are smaller and more manageable, yet capable of achieving the same results as a complete disk clone. This functionality enables various types of forensic analysis to be conducted on the image evidence, including the use of decryption tools that can unlock encrypted content, facilitating thorough investigations.

WinHex is designed for hexadecimal data recovery and enables examiners to perform a variety of operations on data, including hash analysis before and after any modifications. Additionally, tools like History View, Chrome Cache View, and MZView serve as cookie managers, allowing users to examine the details stored within cookies, which can be either saved as cookies.txt or in SQLite format within a single table.

Overall, these tools are invaluable for web forensics, and their application can be expanded with additional tools in future forensic investigations.

\paragraph{\textbf{R3}} Based on the experiments conducted, the preferred mode for detecting potential malicious activities by a suspect is the normal browsing mode. This mode is advantageous because it typically preserves the majority of information related to browsing sessions. However, if the suspect is aware of and utilizes private or portable modes, it becomes more challenging to recover artifacts, although it is still possible to collect key information using forensic tools.

Utilizing memory dump files from the device is another viable option, as they can reveal keywords from search queries and URLs. However, establishing the exact timing of these activities may pose a challenge.

Throughout the evaluation, each browser was assessed using various methodology scenarios where identical web activities were performed. The table below summarizes the information obtained across normal, private, and portable modes for the web browsers tested.

In private browsing mode, cookies expire when the browser is closed; however, if the browser remains open, these cookies can be copied. Bookmarks are accessible in the browser's bookmarks section since they are stored in the root default folder. Even after closing the browser, bookmarks remain available unless manually deleted. While users can clear their download history within the browser, the actual downloaded files still exist in the default downloads folder or in any custom location specified during the download process.

\begin{figure}[!htb]
    \centering
    \includegraphics[width=0.9\linewidth]{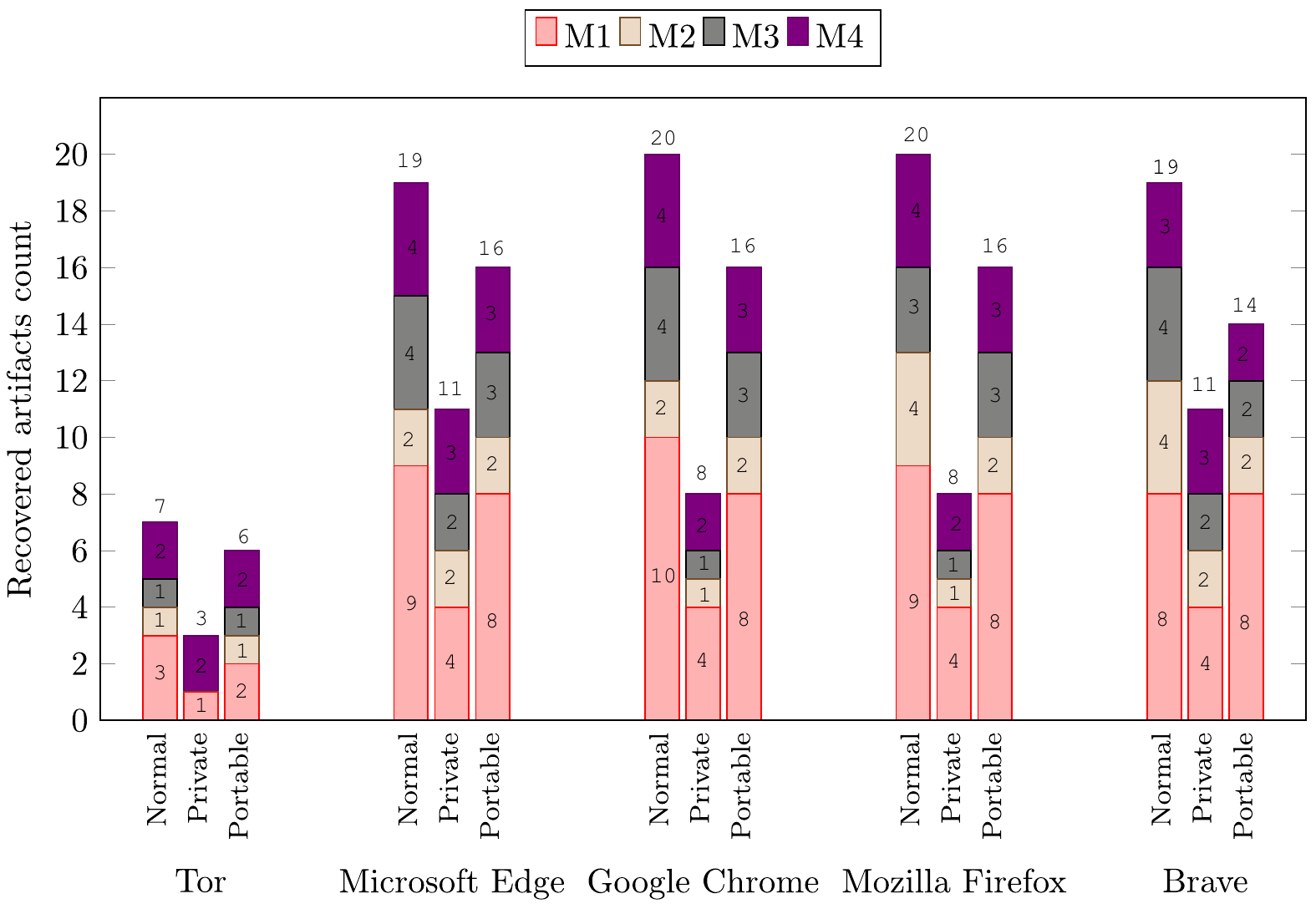}
    \caption{Comparison of four browsers in terms of the number of artifacts recovered across the three browsing modes}
    \label{fig:browser_artifacts_count}
\end{figure}

Table~\ref{tab:artifacts} reports the recovered browser artifacts across four browsing scenarios (M1, M2, M3, and M4) in three browsing modes (normal, private, and portable) for five browsers: Tor, Microsoft Edge, Google Chrome, Mozilla Firefox, and Brave. Figure~\ref{fig:browser_artifacts_count} presents a comparison of four browsers in terms of the number of artifacts recovered across the three browsing modes.
\begin{table}[htb!]
\centering
\caption{Recovered browser artifacts across four browsing scenarios (M1, M2, M3, and M4), in three browsing modes (normal, private, and portable), for five browsers (Tor, Microsoft Edge, Google Chrome, Mozilla Firefox, and Brave)}\label{tab:artifacts}
\resizebox{1\textwidth}{!}{
\begin{tabular}{@{\extracolsep{4pt}}clcccccccccccccccccccc}
\hline
\multirow{2}{*}{Mode} & \multirow{2}{*}{Artifacts} & \multicolumn{4}{c}{Brave Browser} & \multicolumn{4}{c}{Mozilla Firefox} & \multicolumn{4}{c}{Google Chrome} & \multicolumn{4}{c}{Microsoft Edge} & \multicolumn{4}{c}{TOR Browser} \\
\cline{3-6}\cline{7-10}\cline{11-14}\cline{15-18}\cline{19-22}
& & M1 & M2 & M3 & M4 & M1 & M2 & M3 & M4 & M1 & M2 & M3 & M4 & M1 & M2 & M3 & M4 & M1 & M2 & M3 & M4 \\
\hline
\multirow{9}{*}{\rotatebox[origin=c]{90}{Normal Mode}} 
& History & \checkmark & & & & \checkmark & & & & \checkmark & & & & \checkmark & & & & \checkmark & \\

& Bookmarks & \checkmark & \checkmark & \checkmark & \checkmark & \checkmark & \checkmark & \checkmark & \checkmark & \checkmark & \checkmark & \checkmark & \checkmark & \checkmark & \checkmark & \checkmark & \checkmark & \checkmark & \checkmark \\

& Web Cache & \checkmark & \checkmark & \checkmark & \checkmark & \checkmark & & & & \checkmark & & & & \checkmark & & & & & & \\

& Cookies & \checkmark & \checkmark & & & \checkmark & \checkmark & & \checkmark & \checkmark & & \checkmark & \checkmark & \checkmark & & \checkmark & \checkmark & & & \checkmark & \checkmark \\

& Login & \checkmark & & & & \checkmark & & & & \checkmark & & & & \checkmark & & & & \\

& Favicons & & & & & & & & & \checkmark & & & & & & & \\

& Thumbnails & & & & & \checkmark & \checkmark & \checkmark & \checkmark & \checkmark & & \checkmark & \checkmark & \checkmark & & \checkmark & \checkmark & & \\

& Downloads & \checkmark & \checkmark & \checkmark & \checkmark & \checkmark & \checkmark & \checkmark & \checkmark & \checkmark & \checkmark & \checkmark & \checkmark & \checkmark & \checkmark & \checkmark & \checkmark &  \\

& Search Keyword & \checkmark & & & & \checkmark & & & & \checkmark & & & & \checkmark & & & & & & & \checkmark \\

& Top Sites & \checkmark & & & & \checkmark & & & & \checkmark & & & & \checkmark & & & & \checkmark & \\
\hline
\multirow{8}{*}{\rotatebox[origin=c]{90}{Private Mode}}
& History & \checkmark & & & & & & & & & & & & & & & \\

& Bookmarks & & & & & & & & & & & & & & & & \\

& Web Cache & & & & \checkmark & & & & \checkmark & & & & & & & & & & \\

& Cookies & & \checkmark & \checkmark & \checkmark & \checkmark & & & & \checkmark & & & & \checkmark & \checkmark & \checkmark & \checkmark & & & & \checkmark\\

& Login & \checkmark & & & & \checkmark & & & & \checkmark & & & & \checkmark & & & & \\

& Favicons & & & & & & & & & \\

& Thumbnails & & & & & & & & & \\

& Downloads & \checkmark & \checkmark & \checkmark & \checkmark & \checkmark & \checkmark & \checkmark & \checkmark & \checkmark & \checkmark & \checkmark & \checkmark & \checkmark & \checkmark & \checkmark & \checkmark & & \\

& Search Keyword & \checkmark & & & & \checkmark & & & & \checkmark & & \checkmark & \checkmark & & & & \checkmark & \checkmark & & & \checkmark \\

& Top Sites & & & & & & & & \\
\hline
\multirow{8}{*}{\rotatebox[origin=c]{90}{Portable Mode}}
& History & \checkmark & & & & \checkmark & & & & \checkmark & & & & \checkmark & & & & \checkmark & \\

& Bookmarks & \checkmark & & & & \checkmark & \checkmark & \checkmark & \checkmark & \checkmark & \checkmark & \checkmark & \checkmark & \checkmark & \checkmark & \checkmark & \checkmark & \checkmark & \checkmark  \\

& Web Cache & \checkmark & & & & \checkmark & & & & \checkmark & & & & \checkmark & & & & & & \\

& Cookies & \checkmark & \checkmark & \checkmark & \checkmark & \checkmark & & \checkmark & \checkmark & \checkmark & & \checkmark & \checkmark & \checkmark &  & \checkmark & \checkmark & & & \checkmark &\checkmark \\

& Login & \checkmark & & & & \checkmark & & & & \checkmark & & & & \checkmark & & & & & & & \\

& Favicons & & & & & \\

& Thumbnails & & & & & \\

& Downloads & \checkmark & \checkmark & \checkmark & \checkmark & \checkmark & \checkmark & \checkmark & \checkmark & \checkmark & \checkmark & \checkmark & \checkmark & \checkmark & \checkmark & \checkmark & \checkmark & & \\

& Search Keyword & \checkmark & & & & \checkmark & & & & \checkmark & & & & \checkmark & & & & \checkmark & & & \checkmark \\

& Top Sites & \checkmark & & & & \checkmark & & & & \checkmark & & & & \checkmark & & & & & & & \\
\hline
\end{tabular}
}
\end{table}

Normal browsing mode yielded the highest number of browser artifacts across all tested browsers. These artifacts were located in each browser's respective directory and captured using various forensic tools, including SQLite browser extension, SQLite DB Manager, FTK Imager, WinHex, BrowsingHistoryView, ChromeCacheView, InternetEvidenceFinder, MZCacheView, and MZCookies.
While suspects may utilize private browsing mode to disrupt or eliminate their browsing history, significant artifacts often remain in local storage, on the hard disk, and within memory. These remnants are invaluable for investigators, as they help reconstruct user activities and track the sites visited, thereby enhancing the overall forensic analysis. 

In normal browsing mode, a forensic investigation was conducted for each browsing scenario M1 to M4. The analysis revealed that all browsers contained specific artifacts, as detailed in Figure \ref{fig:4}. Scenario M1 produced the highest number of artifacts since no data was cleared after the browsing session, resulting in a wealth of valuable information. In contrast, scenarios M2, M3, and M4 involved clearing history and memory dumps at each stage. Despite these actions, forensic tools still uncovered a total number of artifacts in each browser for each methodology. Notably, Google Chrome yielded the most artifacts, while the TOR browser provided the least.
Similarly, in private browsing mode, each browser retained identifiable artifacts.

In portable mode, more artifacts were recovered compared to private mode. This is because all information related to web activity is stored in the designated folder of the portable web browser. However, if the portable browser is used from a removable device, it may contain fewer artifacts since the entire browser would be removed along with its data. Despite this, information can still be captured in the memory of the host machine, allowing for the recovery of keywords and URLs from searches. This recovered data can be instrumental in reconstructing the suspect's search history.

All browsers store user information and web activity, regardless of whether they are in normal, private, or portable mode. However, portable mode may contain less data because unplugging the portable device can result in the removal of all stored information. Nevertheless, the device's memory dump may still retain valuable information, such as search keywords and URLs.

The artifact recovery analysis across various browsing modes (M1, M2, M3, and M4) provides critical insights into the balance between user privacy and forensic utility. As shown in Table~\ref{tab:artifacts}, browsers such as Microsoft Edge, Google Chrome, and Mozilla Firefox in normal mode (M1) retained a vast amount of artifacts, including history, bookmarks, cookies, and search keywords, making these browsers particularly useful for forensic investigations. Figure \ref{fig:browser_artifacts_count} reports that Google Chrome recovered ten artifacts in M1, followed closely by Microsoft Edge and Mozilla Firefox with nine artifacts. These numbers demonstrate the extensive data retained by these browsers in normal mode, offering valuable forensic evidence. In contrast, TOR Browser, which prioritizes user privacy, yielded only three artifacts in M1, making it the most secure for privacy-conscious users. However, the limited artifact recovery in TOR poses a challenge for investigators who rely on browser data to reconstruct user behavior and activities.

In private browsing mode (M4), the total recoverable artifacts decreased across all browsers, but Microsoft Edge and Brave still displayed relatively high artifact retention, with three artifacts recovered, as shown in Figure \ref{fig:browser_artifacts_count}. Google Chrome's artifact count dropped to 2 artifacts in M4, but it still retained traces of activity, such as downloads and session data, which compromises the perceived privacy of private mode. TOR Browser, on the other hand, recovered only two artifacts in private mode, maintaining its reputation as the most privacy-oriented browser. Despite this, it is evident that no browser completely eliminates traces of activity, even in private mode, with memory dumps and cached data offering some recoverable artifacts. This aligns with the results in Table~\ref{tab:artifacts}, which reveal that even private modes leave certain digital traces that forensic investigators can analyze.

The findings emphasize that the efficacy of browsers varies based on the mode used and the context of usage. From a forensic perspective, normal mode (M1) in browsers like Chrome and Edge is the most beneficial, as it provides the richest set of artifacts for analysis. Private mode (M4), although marketed for privacy, still leaves some recoverable data, with Microsoft Edge retaining more artifacts than expected. TOR Browser emerges as the most privacy-centric option, though its limited recoverability could be a drawback in forensic investigations. As noted in the results, while TOR's artifact recovery is minimal, forensic investigators can still extract small traces, especially in memory dumps. Therefore, while no browser completely fulfills the promise of privacy or security, the trade-off between privacy and forensic recovery must be carefully navigated, and users should remain aware of the potential limitations of private modes, as demonstrated by the differences in artifact recovery.

\section{Challenges and Opportunities}\label{sec:challenges}

Web browser forensics presents various challenges due to the dynamic nature of browser technologies and the evolving landscape of user privacy and security. While this research reviewed several tools and techniques for extracting digital artifacts from browsers, the limitations and challenges encountered highlight the need for further exploration and development in this field. Below, we identify key challenges and outline directions for future research.

\paragraph{Detection of browser file directories}
One of the primary challenges in browser forensics is the detection of browser file directories. Tools, such as DB Browser for SQLite and MZCookiesView, currently lack the capability to automatically identify the default locations of these directories, which necessitates manual intervention \citep{1}. This manual process can result in inefficiencies and potential inaccuracies in the forensic analysis \citep{4}. Furthermore, it has been observed that numerous existing forensic tools do not support the automatic detection of browser directories. To address this issue, researchers should focus on enhancing forensic tools to automatically detect and locate browser directories across various operating systems. This improvement would reduce the need for manual configuration and enhance the accuracy of data collection.

\paragraph{Operating system environments}
The forensic tools and browser tests conducted in this research were limited to the Windows operating system environment. Previous studies have utilized various platforms, including Android \citep{4,13} and Linux \citep{6,19}, when performing digital forensics. This specific focus poses a challenge, as different operating systems may manage browser data in distinct ways. To enhance the robustness of digital forensic investigations, future research should expand to include additional environments, such as macOS, Linux, and mobile operating systems. This broader approach will provide a more comprehensive understanding of how browser data can be retrieved across different platforms.

\paragraph{Anti-forensics techniques and browser extensions}
Anti-forensics, or counter-forensics, poses a significant challenge in the field of web browser forensics. Techniques such as data deletion, overwriting, steganography, onion routing, and timestamp alteration are employed to obscure or destroy digital evidence. These techniques make it difficult for investigators to collect and analyze browser data. Tor browser was used for retrieving artifacts for forensic evidence \citep{3,7,9}, where an extra layer of onion routing is added. Moreover, browser extensions, designed to enhance user privacy and security, can complicate forensic investigations by encrypting browsing history and hiding user data. These extensions may also interfere with metadata, making it challenging to gather accurate information. In the future, researchers should focus on developing robust countermeasures against these anti-forensics techniques, ensuring that critical evidence can still be recovered even in the face of such tactics and should investigate techniques to bypass or mitigate the effects of such extensions, enabling investigators to access the data necessary for their work.

\paragraph{Lack of standardization for portable web browsers}
Portable web browsers, which can be run from external storage devices without installation on a host machine, present unique challenges in forensic investigations. These browsers store user data within the portable device \citep{3,10}, making it difficult to acquire and analyze the data if the device is not available. Future research could focus on establishing standards for forensic analysis of portable browsers and developing tools that can effectively retrieve data from these devices.

\section{Conclusion}\label{sec:conclusion}
The recovery of artifacts during browser forensics is essential for effective investigations. The core findings of this extensive research provide a detailed analysis of browser forensic tools and techniques. We conducted a comprehensive analysis of various browsers and proposed a methodology for investigators to select the most suitable forensic tools based on specific needs. Each tool has its strengths and weaknesses that must be considered in different scenarios. Investigators can use our research as a guide to compare their current toolkits with others and potentially improve their forensic tools. Currently, no browser fully provides the privacy and anonymity it promises, including private mode. It is evident that all browsers store user browsing session and activity data, even in private mode, where many users may believe their history is not being captured. Information such as search keywords and URL visits are stored in memory that can be retrieved by the forensics investigators.

Future research should focus on other browsers in the market, particularly those based on different operating systems, to understand where and how artifacts are stored and accessed in various scenarios. While our study concentrated on popular browsers, updates are periodically rolled out to enhance user privacy and security. However, in terms of forensics, these browsers still store user information based on online activities.

\section*{CRediT authorship contribution statement}
\textbf{Rishal Ravikesh Chand:} Conceptualization, Methodology, Visualization, Formal analysis, Writing – original draft, Writing – review \& editing. 
\textbf{Neeraj Anand Sharma:} Conceptualization, Methodology, Validation, Writing – review \& editing.
\textbf{Ashad Kabir:} Validation, Visualization, Supervision, Writing – review \& editing.

\section*{Declaration of competing interest}
The authors declare that they have no known competing financial interests or personal relationships that could have appeared to influence the work reported in this paper.

\section*{Funding}
This research did not receive any specific grant from funding agencies in the public, commercial, or not-for-profit sectors.

\section*{Data availability}
No data was used for the research described in the article.
\bibliographystyle{elsarticle-harv}
\bibliography{reference}

\end{document}